\documentclass[aps,prx,reprint,superscriptaddress,longbibliography]{revtex4-2}

\usepackage{hyperref}
\usepackage[english]{babel}
\usepackage[utf8]{inputenc}

\usepackage{amsmath}
\usepackage{amssymb}
\usepackage{mathtools}
\usepackage{mathrsfs}

\usepackage{graphicx}
\usepackage{xcolor}

\usepackage{float}
\usepackage{quantikz}
\usepackage[colorinlistoftodos, color=green!40, prependcaption]{todonotes}

\usepackage{algorithm}
\usepackage{algorithmicx}
\usepackage{algpseudocode}

\usepackage{cleveref}
\usepackage{balance}

\usepackage{physics}
\usepackage{csquotes}
\usepackage[T1]{fontenc}

\begin{document}
\title{Suppressing Frequency Collisions in a Quantum Processor}
\author{Kuljeet Kaur}
\email[Correspondence email address: ]{k.kaur@fz-juelich.de}
\author{Xuexin Xu}
\author{Mohammad H. Ansari}%
    \affiliation{Peter Gr\"unberg Institute 12 , Forschungszentrum J\"ulich, J\"ulich 52428, Germany}

\preprint{AAPM/123-QED}
\begin{abstract}
Frequency crowding, the close spacing of qubit transitions, limits superconducting quantum processors through crosstalk, leakage, and spectator-dependent errors. We introduce a Hamiltonian-based framework that maps two- and three-body collision mechanisms onto a Max-k-Cut problem on qubit frequency allocation. In order to make larger lattices tractable, we introduce a cluster-stitching protocol that
decomposes larger lattices into overlapping subgraphs and reconciles their local frequency assignments into globally consistent configurations. We tested our proposed framework across multiple qubit architectures to suppress dominant collision penalties by revealing a hardware residual crosstalk. We benchmark digitized counterdiabatic QAOA (DC-QAOA), standard QAOA, and simulated annealing. Counterdiabatic terms improve QAOA optimization near steep collision penalties, while simulated annealing achieves the lowest costs on the tested instances. These results indicate that our hardware-aware optimization connects microscopic interaction physics to processor-level frequency planning. These results enable a systematic solver for benchmarking of scalable software–hardware co-design.
\end{abstract}
\maketitle 
\section{Introduction}
Analog quantum computation has provided a powerful source of inspiration for condensed-matter physics over the past two decades. It has stimulated intense experimental and theoretical activity around quantum processors and communication devices, particularly their unusual transport properties and potential to preserve quantum coherence and entanglement over extended timescales  \cite{Daley2022,Pazem2025Brainbox}.  An analog quantum simulator is a controllable quantum device engineered such that its natural Hamiltonian closely reproduces that of a quantum system of interest. Rather than executing a long sequence of universal logic gates, the simulator prepares an initial state, allows it to evolve continuously under designed interactions, and measures observables such as correlations, transport, excitation dynamics, and order parameters \cite{Nori2014}. In this approach, the relevant quantum states and dynamics emerge directly from continuous evolution under an engineered Hamiltonian. Its principal strength is direct experimental access to many-body dynamics and quantum phases that are often difficult to calculate classically. 

One of the most advanced forms of analogue quantum simulation can be realized in superconducting quantum processors. These processors consist of artificial atoms, each implemented as a superconducting electrical circuit operating as a qubit. At millikelvin temperatures, electrical current flows without ordinary resistance, allowing the circuit energy levels to exhibit quantum behavior \cite{Blais2021CircuitQED,Krantz2019QuantumEngineer,Kjaergaard2020SuperconductingQubits}.  Two-dimensional architectures have been tested to demonstrate high-fidelity state preparation and gate operations, as well as quantum computational advantage in carefully defined sampling tasks \cite{101038s41586-019-1666-52019NatureAruteMartinisQuantumsupremacyusingaprogrammablesuperconductingprocessor,collaborators2025quantum-26f}.  Large-scale gate-based processors containing hundreds of qubits have been pursued, but their performance is limited by two distinct forms of variation \cite{Blais-RevModPhys2021,ansari2011noise-ac0}: Firstly, variations in fabrication and design parameters cause nominally identical qubits to acquire different frequencies, anharmonicities, couplings, and coherence properties, leading to behavior that deviates from the intended design \cite{Mohseni2024};  Secondly, variations arise during operation: gate fidelities, residual interactions between qubits, and idle-state behavior can differ substantially from their designed values \cite{XuAnsari2021ZZFreedom,Ku2020SuppressionZZ,Xu2026SurfaceCodeHamiltonian}. 
 While superconducting circuits can be made relatively insensitive to weak or moderate disorder, and non-equilibrium quasi-particles can be mitigated through engineered traps \cite{[Pan2022],PhysRevB.91.195434}, the operation of novel qubits in large-scale processors is still seriously hindered by disorder \cite{VanDamme2024}, crosstalk and state-preparation-and-measurement (SPAM) \cite{Hangleiter2024}, noise and correlated relaxation errors \cite{Wilen2021}. 
 
 Scaling up the number of qubits is fundamentally bottlenecked by frequency crowding, in which unavoidable fabrication variances create a highly fragile spectral landscape. Increasing qubit connectivity further constrains this allocation: each frequency choice affects collision conditions involving multiple neighboring qubits. Both engineered couplings and parasitic capacitive interactions contribute to these constraints, affecting gate and idle fidelities through frequency collisions, leakage, and spectator effects, \cite{Morvan2022OptimizingSCQP,Zhang2025FrequencyAllocation,Hertzberg2021LaserAnnealing,101103physrevb1000245092019PhysicalReviewBAnsariAnsariSuperconductingqubitsbeyondthedispersiveregime3za}. 
 
Motivated by these challenges, we develop a Hamiltonian-informed method for assigning frequencies to a lattice of superconducting qubits. Accurate allocation requires accounting for interactions beyond isolated qubit pairs, as the surrounding lattice can substantially modify gate dynamics. Three-qubit lattice Hamiltonian \cite{Xu2024LatticeHamiltonians} and their recent extension to five-qubit surface Hamiltonian \cite{Xu2026SurfaceCodeHamiltonian} reveal regimes in which effective three-body interactions exceed two-body terms, reversing the expected interaction hierarchy. These findings motivate incorporating multi-qubit effects into frequency-allocation constraints. \\

By optimizing the frequency configuration, our approach aims to reduce spectral crowding and unwanted crosstalk while preserving the resonance or detuning conditions required for gate operations. A principled frequency allocation thus supports high-fidelity two-qubit gates at scale, by placing qubits away from leakage transitions while preserving the resonance and detuning conditions the gates themselves require, the framework provides a route toward low-crosstalk operation in larger processors \cite{Chow2011CrossResonance,Sheldon2016CrosstalkTuning}. As processors grow, additional constraints on frequency placement and unwanted interactions make systematic allocation increasingly important \cite{Yan2018TunableCoupler,Mundada2019CrosstalkSuppression}. Existing approaches include discrete heuristics such as graph coloring and stochastic methods such as simulated annealing and Monte Carlo search. Discrete formulations also allow frequency allocation to be addressed using variational quantum algorithms.

The Quantum Approximate Optimization Algorithm (QAOA) \cite{Zhou2020QAOA} provides a variational framework for combinatorial optimization on near-term quantum hardware \cite{Preskill2018NISQ}. Its performance depends on both parameter optimization and the fidelity of the implemented unitaries \cite{Zhou2020QAOAPerformance,Cerezo2021VariationalQuantumAlgorithms}. In superconducting processors, frequency placement influences collisions and the effects of parasitic couplings and crosstalk during execution \cite{Harrigan2021QAOANonPlanar,Arute2019QuantumSupremacy}. These hardware errors can degrade the prepared states and achievable solution quality \cite{Guerreschi2019QAOAHundredsQubits}, motivating the joint design of optimization algorithms and hardware frequency configurations.

We develop a hardware-aware framework for frequency allocation in superconducting processors. We encode two-body and three-body collision mechanisms, including neighboring-qubit collisions, gate-induced leakage, and spectator effects, into a discrete cost Hamiltonian derived from circuit physics. We formulate the allocation problem as a Max-\(k\)-Cut-inspired task  \cite{Fuchs2021MaxKCut} and compare DC-QAOA, standard QAOA \cite{Farhi2014QAOA}, and classical simulated annealing \cite{Kirkpatrick1983}. To address larger processors, we introduce a cluster-stitching protocol that combines solutions on overlapping subgraphs into a globally consistent frequency assignment. The framework provides a physically grounded approach to collision mitigation and a common benchmark for classical and quantum optimization.

\begin{figure}[htbp]

\hspace*{-7.5cm}\text{(a)} \\
\includegraphics[width=0.48\textwidth]{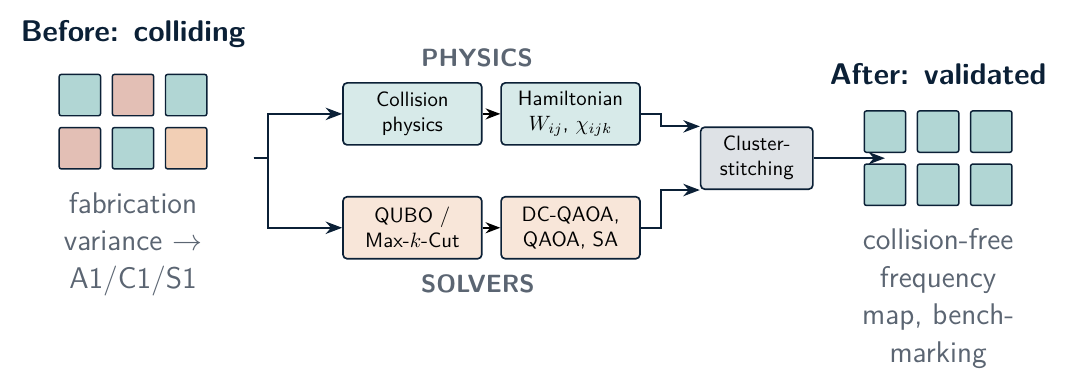}\\ 
\hspace*{-7.5cm}\text{(b)} \\
\includegraphics[width=0.45\textwidth]{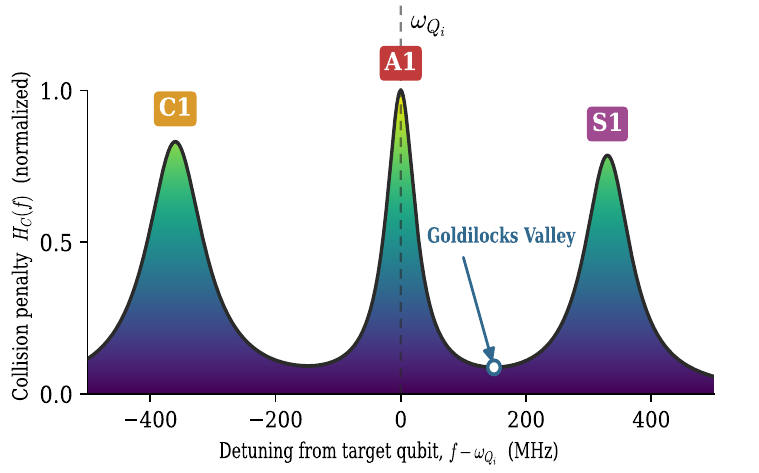} 
\caption{(a) Overview of the frequency-allocation framework for six qubits, represented by boxes arranged in two rows. Left: the initial assignment contains three qubits involved in frequency collisions (red), while the remaining qubits are collision-free (green). Center: Perturbative estimation of the collision-Hamiltonian coefficients $\mathcal{W}_{ij}$ and $\chi_{ijk}$, defining the cost function optimized using simulated annealing, QAOA, and DC-QAOA. Cluster stitching reconciles overlapping local solutions into a globally consistent assignment. Right: the resulting collision-free frequency map, with optimization performance benchmarked against classical simulated annealing. (b) Frequency-collision landscape for two coupled transmon qubits, showing exclusion zones around the target qubit frequency $\omega_{Q_i}$.Collision conditions are labeled following the taxonomy of Ref. \cite{Chow2011CrossResonance}: the letter denotes the physical mechanism ($\mathtt{A}$: always-on, $\mathtt{C}$: control-drive, $\mathtt{S}$: spectator, $\mathtt{E}$: entangling-drive), and the number denotes the transition order  $'\mathtt{1}'$ for the primary, first-order instance (typically involving the $(\ket{0}\rightarrow\ket{1})$, and $'\mathtt{2}'$ for the higher-order, anharmonicity-shifted instance (typically involving the $(\ket{1}\rightarrow \ket{2}$ leakage transition, offset by 
$\eta)$. In the frequency-spacing regime considered here, the leading terms $\mathtt{A1}$, $\mathtt{C1}$, and $\mathtt{S1}$ dominate the collision landscape, and the higher-order conditions $\mathtt{A2}$, $\mathtt{E2}$ are suppressed and can be neglected (see Appendix \ref{app.A} for the full taxonomy).}
\label{fig:big_wide_fig}
\end{figure}

The paper is organized as follows. Section II introduces the physical framework and collision mechanisms. Section~\ref{sec:method} constructs the discrete cost Hamiltonian and its quadratic unconstrained binary optimization~(QUBO) and Ising representations, while Section IV derives the two-body and three-body coefficients from the coupled-transmon model. Section V presents DC-QAOA and cluster stitching. Section VI describes the classical baseline and compares the three optimization methods across processor geometries. Section VII discusses limitations and future directions.



\section{Method}
\label{sec:method}
High-fidelity multi-qubit operations in superconducting architectures require frequency configurations that accommodate intended interactions while avoiding unwanted resonances within the available spectral bandwidth \cite{Devoret2013SuperconductingCircuits,Wendin2017QuantumInformationReview}. 
These constraints apply to both fixed-frequency and tunable qubits. Multi-mode couplers provide additional ways to engineer qubit interactions, as explored in a coupled two-mode design \cite{jiang2026} and an experimental left-handed meta-material ring resonator \cite{PRXQuantum.5.020325}. Fabrication-induced variations and operational frequency settings can bring coupled qubits, driven transitions, or spectator qubits close to unwanted resonances. We consider three collision mechanisms relevant to frequency allocation: $\mathtt{A1}$, the interacting nearest-neighbor collisions, ($\mathtt{C1}$) gate-induced collisions, and ($\mathtt{S1}$) spectator-qubit-induced collisions. The complete classification of the collision conditions considered here is summarized in Appendix (\ref{app.A}). 

\begin{description} 
\item[A1-neighboring collisions] occur when the fundamental transition frequencies of adjacent qubits become nearly degenerate, $f_i\approx f_j$ for $i,j$ being nearest neighbor qubit labels, say $|i-j|=1$. In coupled qubits, this condition enhances unwanted $ZZ$ interaction, which causes an always-on coherent error, no matter whether the gate is on or off \cite{Ku2020SuppressionZZ,XuAnsari2021ZZFreedom}. 
\item[C1-Gate clashes]Gate clashes can arise when drives used for two-qubit gate operations approach unintended non-computational transitions of nearby qubits. The releva,nt drive whether and leakage channels depend on the gate architecture; e.g., microwave-driven gates and parametrically activated gates can couple to different leakage transitions. In this work, we focus on microwave-driven cross-resonance(CR) gates, for which the drive frequency is typically set to the target-qubit frequency, $f_{\rm drive}=f_j$. A clash occurs when this drive approaches a non-computational transition of a neighboring qubit, such as the $\ket{1}\rightarrow\ket{2}$ transition of a transmon at $f_i+\eta_i$, potentially inducing leakage from the computational subspace.

\item[S1 spectator effects] arise when a spectator qubit with frequency $f_k$ modifies the effective interactions or transition energies of an active target--control pair $(f_i,f_j)$. These effects can introduce spectator-state-dependent shifts and unwanted multi-qubit interactions, reducing entangling-gate fidelity. \end{description}

To treat frequency allocation as a discrete optimization problem solvable by a variational quantum algorithm, the physical collision mechanisms must be mapped onto a diagonal cost function, $H^{\mathtt{C1}}$. We construct a multi-tiered penalty system where the total objective function is decomposed into distinct energy contributions. 

For a neighboring qubit pair $(i,j)$, an A1 collision is represented as $P^{\mathtt{A1}} =\Theta(\delta^{\mathtt{A1}} -|f_{ij}|)$,
with $\delta^{\mathtt{A1}}$ being the minimum allowed frequency separation and $\Theta$ is the Heaviside step function. The frequency detuning is $f_{ij}=f_i-f_j$. The  contribution of A1 collisions on the cost function is 
\begin{equation}
\label{eq.HA1}
    H^{\mathtt{A1}} =\sum_{(i,j)\in \mathbb{E}} \mathcal{W}_{ij} \Theta(\delta^{\mathtt{A1}} -|f_i-f_j|),
\end{equation}
in which the set of coupled qubit pairs being $\mathbb{E}$, and the weights associated with each pair $(i,j)$ being $\mathcal{W}_{ij}$. Similarly, the other collision contributions can be written as
\begin{equation}
\label{eq.HC1}
    H^{\mathtt{gate}} =\sum_{(i,j)\in \mathbb{E}} \mathcal{W}_{ij} \Theta(\delta_{\mathtt{gate}} -|f_j-(f_i+\eta_i)|),
\end{equation}
where $\eta_i$ being the anharmonicity of qubit $i$, and $\delta_{C1}$ determines resonance between $0\to 1$ transition frequency in qubit $i$ and $1\to 2$ transition in the qubit $j$, which corresponds to the leakage to non-computational states in at least one of the qubits. 

The spectator contribution to the cost function consist of three qubits two of which are under a gate interaction and the third one is spectator and only impacted indirectly by the gate; which we will denote as 
\begin{equation} 
\label{eq.HS1}
    H^{\mathtt{S1}} =\sum_{(i,j,k)\in \mathbb{T}} \chi_{ijk} \Theta(\delta_{S1} -|f_k-{f_{ij}^{\rm{g}}}|).
\end{equation}
where $\mathbb{T}$ denotes the relevant set of target control spectator triplets $f_{ij}^{\rm{g}}$ is the frequency difference associated with the active gate on qubit $i$ and $j$, and $\chi_{ijk}$ being the cost function weight corresponding to the qubits $i$, and $j$ under the gate and $k$ spectator.

If only collision penalties are considered, the optimization may favor excessively large frequency separations. However, large detuning can also reduce gate (temporal) length $T_g$. To balance spectral separation and gate performance, we introduce the gate-time contribution to the cost function. For the gate interaction between the two qubits $i$ and $j$ being $J_{ij}$ , the characteristic gate time scales  $f_{ij}/J_{ij}^2$. By defining the desired gate length to be  $T_g^\text{target}$, we can write
\begin{equation}
\label{eq.HTg}
    H^{\mathtt{Tg}} = \sum_{(i,j)\in \mathbb{E}} \alpha_{ij} \left(\frac{T_g(i,j)-T_g^\text{target}}{T_g^\text{target}}\right)^2
\end{equation}
with the co-factor \(\alpha_{ij}\) determining the relative importance of the gate-time constraint between the two pairs of qubits.

Summing all of these cost functions in Eqs. (\Cref{eq.HA1,eq.HC1,eq.HS1,eq.HTg}) define the full cost function 
\begin{equation}
\label{eq. HCost}
H^{\mathtt{C}}=H^{\mathtt{A1}}+H^{\mathtt{gate}}+H^{\mathtt{S1}}+ H^{\mathtt{Tg}}
\end{equation}. 

{\bf QUBO mapping:} \label{sec:discretization} QUBO mapping reformulates an optimization problem as minimizing a quadratic function of binary variables, \(E(\mathbf{x})=\mathbf{x}^{T}Q\mathbf{x}\), where \(x_i\in\{0,1\}\).  For implementing a discrete optimization algorithm, we restrict the frequencies to be limited to a discrete set of $M$ numbers 
\begin{equation}
\label{eq.FM}
    \mathbb{F}^M=\{f^{(1)},f^{(2)},...,f^{(M)}\}.
\end{equation} 
with $f^{(m)}$ being a discrete frequency in between $f^{(1)}$ and $f^{(M)}$, for $1\leq m\leq M$.  We introduce the binary variables
\begin{equation}
    x_{i,m} = \begin{cases} 
          1 & \text{qubit}~i~\text{assigned frequency}~f^{(m)}\\
          0 & \text{otherwise}       
       \end{cases}
\end{equation}
with one strict constraint $\sum_m x_{i,m}=1$. This constraint warrants that only one frequency is assigned to each qubit. The frequency of qubit $i$ can be written as  $f_i = \sum_m f^{(m)}x_{i,m}$. 

In a qubit lattice, the interaction between qubit $i$ and $j$ can be represented as an edge that connects the two vertices at the positions $(i,j)$. If the two qubits are in resonance, this means that the frequency difference, say $f^{(m)} - f^{(n)}$, is smaller than the tolerance domain of $\delta^{\mathtt{A1}}$. This brings in a penalty that can be contributed to the frequencies $(f^{(m)},f^{(n})$ as values. Putting all these together, we can define the penalty associated with the frequency values as 
\begin{equation}
\label{eq.edge_penality}
C_{mn}^{\mathtt{\mathtt{A1}}}=\begin{cases}
    1,\qquad |f^{(m)}-f^{(n)}|< \delta^{\mathtt{A1}}\\
    0, \qquad \text{otherwise}
\end{cases}.
\end{equation}

Summing up over all interacting edges will define the cost of neighboring frequency overlap to become 
\begin{equation}
    H^{\mathtt{A1}}= \sum_{(i,j)\in \mathbb{E}} \quad \sum_{m,n \in \{1,\cdots,M \}} \mathcal{W}_{ij}C_{mn}^{\mathtt{\mathtt{A1}}}x_{i,m}x_{j,n}\label{QUBO}
\end{equation}
This represents the collision in QUBO formalism. Further, we write the above Hamiltonian in terms of Pauli operators. Using the following definition for $z_{i,m} \equiv (1-2x_{i,m})$,~which assigns $1\ (-1)$ to $z$ where $x=0\ (1)$.  Substituting this in Eq.\eqref{QUBO} make it 
$H^{\mathtt{A1}} =E_0 +\sum_{i,m}h_{i}(f^{(m)}Z_{i,m} +\sum_{i,j}W_{ij}^{(m,n)}Z_{i,m}Z_{m,n}$. 
Following the similar definition and promoting the $Z$ into an all-finite-value vector of frequencies assigned to qubits, the full cost function of Eq. (\ref{eq. HCost}) in binary form will become
\begin{equation} 
\label{eq. Hc}
	H^{\mathtt{C}} = \sum_{i,j}\mathcal{W}_{ij}\hat{Z}_i\hat{Z}_j+\sum_{ijk}\chi_{ijk}\hat{Z}_i\hat{Z}_j\hat{Z}_k	+ \sum_k P(f_k)\hat{Z}_k.
\end{equation} 
in which we removed the coefficients $m$ and $n$ as we assign $m=n$ and $m\neq n$ in the coefficient of $ZZ$ and $ZZZ$ terms \cite{Wang2020XYMixers}. Here, the coefficient $ \mathcal{W} _ {ij} $ functions as a heavy penalty weight that activates if the chosen frequencies for adjacent nodes $i$ and $j$ trigger an $\mathtt{A1}$ or $\mathtt{C1}$ condition. 

Similarly, $\chi_ {ijk} $ assigns a penalty based on the joint configuration of a three-qubit cluster, steering our frequency optimization away from situations with spectator qubit frequency due to its near-resonance with another qubit in its vicinity, which destabilizes a neighboring gate. 

The function $P(f_k)$ penalizes assignments that stray outside the physical limits of the hardware fabrication capabilities or the control electronics bandwidth. Crucially, it balances the $T_ g$ constraint, preventing excessive detuning that would drastically reduce the speed and performance of a two-qubit gate. 

\begin{figure}[h]
	\centering
	\includegraphics[width=1.0\linewidth]{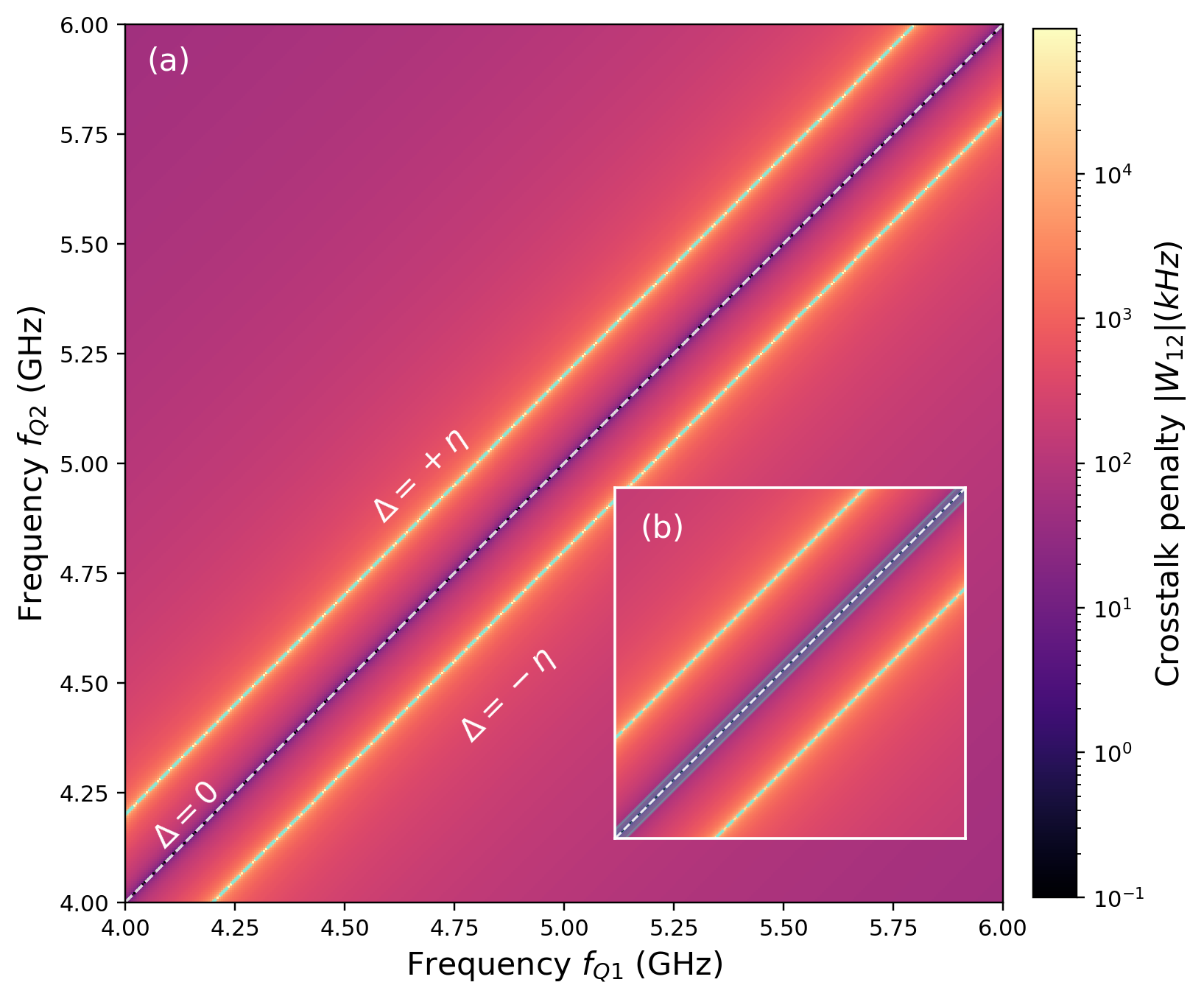}
    \vspace{-0.7 cm}
	\caption{Hardware penalty function $|\mathcal{W}_{12}|$ evaluated as a function of qubit frequencies $f_{Q1}$ and $f_{Q2}$ on a logarithmic color scale. (a) Full tuning range with two bright diagonal bands marks the anharmonicity shifted resonance condition $f_{12} \approx \pm \eta$, while the main diagonal $(f_{12}=0)$ remains dark as the direct dispersive shifts cancel exactly at zero detuning, so the true danger zone is displaced from, not centered on, frequency degeneracy. (b)Inset resolving the narrow safe valley immediately surrounding $f_{12}=0$, bounded on both sides by the $\pm\eta$ resonances.}
	\label{fig:spectralTraffic}
\end{figure}
The mapping of these coupled constraints yields a highly non-convex, rugged energy landscape. As visualized in the 2D hardware penalty landscape parameter spaces, the landscape is defined by steep, divergent "penalty walls" or ridges. These ridges represent forbidden spectral regions where catastrophic collisions reside.
Because the parameters are interconnected through the network topology, a frequency shift in one node inevitably creates a ripple effect across its neighbors, generating an optimization space dense with deceptive local minima. 

 Traditional classical optimization heuristics frequently stall against these steep ridges. The goal of our variational solver is to navigate this rugged terrain to locate the dark, low-energy "Goldilocks Valleys." In these valleys, all primary hard constraints are simultaneously satisfied, reducing the system's energy down to the residual crosstalk floor dictated by the physical limits of the transmon anharmonicity.

{\bf $ZZ$ and $ZZZ$ strengths: } In systems with perturbatively valid effective qubit interactions\cite{Koch2007Transmon}, capacitive or inductive couplings between weakly anharmonic oscillators provide stability of energy eigenvalues with respect to charge dispersion. 

Starting from the full circuit Hamiltonian, a superconducting quantum processor can be decomposed into local clusters of $n$ interacting qubits, namely \emph{$n$-cluster}. Standard circuit QED methods deal with pairwise interactions, which are considered to be $2$-clusters. Ref. \cite{Xu2024LatticeHamiltonians} extends circuit QED to $3$-clusters and recently Ref. \cite{Xu2026SurfaceCodeHamiltonian} generalized the Hamiltonian to three-body interactions in building blocks with five qubits and 9 couplers. 
Perturbatively in the limit of $|J_{ij}| \ll |f_{ij}|$, 2-clusters and higher-order-clusters can be determined perturbatively. Ref. \cite{Xu2026SurfaceCodeHamiltonian} introduces a diagrammatic approach for estimation of all these terms to any arbitrary perturbative order. The results emerge from virtual transitions through non-computational states \cite{Nigg2012BlackBox}. The two-body $ZZ$ and three-body $ZZZ$ interactions calculated from 5-cluster analysis show the strength of interactions to be \cite{Xu2026SurfaceCodeHamiltonian,Xu2024LatticeHamiltonians}, 
\begin{align}
\label{eq_W_chi}
\mathcal{W}_{ij} &= \gamma_{ij} +\sum_k \gamma_{ijk},\\
\chi_{ijk} &=8\sum _{n=0,1}K_{ijk}(n)  
\end{align}
with these parameters being evaluated in Ref. \cite{Xu2026SurfaceCodeHamiltonian} as follows 
\begin{equation}
    \label{eq. ZZ perturb}
\gamma_{ij}=\frac{\left({J_{ij}^{01}}\right)^2}{f_{ij}^{01}} -\frac{\left({J_{ij}^{10}}\right)^2}{f_{ij}^{10}}~\ \text{and}~ \gamma_{ijk}=4(F_{ijk}-K_{ijk}(1))
\end{equation} 
with $f_{ij}^{mn} \equiv (f_i+m\eta_i)-(f_j+n \eta_j)$ being the frequency difference between the transition $|m\rangle \to |m+1\rangle $ in qubit $i$ and the transition $|n\rangle \to |n+1\rangle $ in the qubit $j$. The three-body correction to the pairwise interaction is in the following two key parameters:  
\begin{eqnarray}
   && F_{ijk} = \sum_{n=0,1}\frac{J_{ij}^{00}J_{ik}^{0n}J_{jk}^{00}}{f_{ik}^{0n}f_{jk}^{0n}}-\sum_{n=0,1}\frac{J_{ij}^{1-n,n}J_{ik}^{1-n,0}J_{jk}^{n0}}{f_{ik}^{1-n,0}f_{jk}^{n,0}} \nonumber \\
   && K_{ijk}(n) = -\sum_{\substack{r,s,t=(i,j,k)\\{r<s,t(\neq s)}}}\frac{J_{rs}^{n,n}J_{rt}^{n,1-n}J_{st}^{n,1-n}}{f_{rt}^{n,1-n}f_{st}^{n,1-n}}
\end{eqnarray}

The effective three-body interactions, quantified by $\chi_{ijk}$, arise at fourth order in perturbation theory and reflect correlated virtual excitation pathways among triplets of qubits. These terms capture processes in which excitation is exchanged through multiple intermediate channels, leading to non-additive corrections beyond pairwise coupling. The resulting expression involves contributions from distinct virtual transition paths, encapsulated in  $F_{ijk}$, which represent constructive and destructive interference between pathways, and $K_{ijk}(n)$, which account for permutation-dependent corrections due to intermediate occupation-dependent energy denominators (See Ref. \cite{Xu2026SurfaceCodeHamiltonian} for diagrammatic formalism and details).

{\bf Algorithmic Implementation:}  
To identify the optimal frequency configuration within the non-convex hardware landscape, we implement a modified Quantum Approximate Optimization Algorithm~(QAOA) with digitized counterdiabatic terms (DC-QAOA). In the standard QAOA implementations, a trial state is prepared by alternately applying a cost unitary layer and a transverse-field mixer layer  $H^{\mathtt{M}}=\sum_{i=1}^{N}\hat{X}_i$, on an initial uniform-superposition  $\ket{+}^{\otimes N}$
However, when dealing with the steep, divergent energy barriers characterizing problems such as the frequency allocation problem, standard mixers struggle to explore the state space effectively. They frequently get trapped in high-energy non-physical local minima \cite{Wurtz2021MaxCutPerformanceGuarantees}. To overcome this fundamental limitation, our ansatz incorporates a two-qubit mixer operation as an additional mixer layer. Here, we incorporate Digitized Counterdiabatic (CD) steering layer ~$(H_{CD})$. The unique inclusion of the $H_{CD}$  operator allows the quantum state vector to closely track the instantaneous ground state of the system, mimicking a non-adiabatic transition across highly rugged potential energy surfaces. \\
For a depth of $p$ variational layers, the parameter-dependent trial state $\ket{\Psi(\gamma,\beta,\alpha)}$ is formally constructed by the sequential application of the following parameterized operators as
\begin{equation}
	\ket{\psi(\boldsymbol{\gamma}, \boldsymbol{\beta},\boldsymbol{\alpha})} = \prod_{i=1}^{p} e^{-i (\beta_i H^{\mathtt{M}}+\alpha_i H_{CD})} e^{-i \gamma_i H^{\mathtt{C}}} \ket{+}^{\otimes n},
\end{equation}
where $\boldsymbol{\gamma} = (\gamma_1, \ldots, \gamma_p)$ and $\boldsymbol{\beta} = (\beta_1, \ldots, \beta_p)$ are variational parameters optimized by a classical optimizer. 

The primary driver of our algorithm's enhanced search capability is the structure of the digitized counterdiabatic mixer, $H_{CD}$. While the standard transverse mixer applies single-qubit bit-flip operations across all states,~the counterdiabatic term introduces correlated multi-qubit driving terms. Following the nested-commutator construction of the adiabatic gauge potential \cite{Chandarana2022DCQAOA}, applied to our Ising-type cost Hamiltonian $H^{\mathtt{C}}$, as shown in the appendix \ref{app.B} the leading-order term takes the form: 
\begin{equation}
	H_{CD} = \sum_{ij}A_{ij} ( \hat{Y}_i\hat{Z}_j + \hat{Y}_j\hat{Z}_i)
\end{equation}
where $ A _ {ij} $ represents the coupling coefficient between adjacent nodes in the chip topology. Physically, this operator couples a Pauli-$\hat{Y}$ rotation on a target qubit \textit{i} directly to the Pauli-$\hat{Z}$ state (which maps to the operational frequency) of its neighbor \textit{j}.~This correlation introduces a form of quantum gradient sensing. When the state vector encounters a steep penalty wall such as a configuration yielding an $\mathtt{A1}$ or $\mathtt{C1}$ collision where the $\hat{Z}$ operator registers the massive energy penalty, and the cross-coupled  $\hat{Y}$ drive actively steers the probability amplitudes away from that forbidden region.~This mechanism allows the algorithm to effectively navigate around high-energy penalty walls rather than being blocked by them, guaranteeing a direct, stable trajectory toward the target low-crosstalk "Goldilocks Valleys."

\begin{figure*}[htbp]
\centering
\begin{center}
\begin{quantikz}
\lstick{$\ket{0}$}&\gate{H} & \ctrl{1} & \qw                 & \ctrl{1} & \gate{R_x(2\beta)} & \gate{R_{CD}(\alpha)} & \meter{} \\
\lstick{$\ket{0}$}& \gate{H} & \targ{}  & \gate{R_z(2\gamma)} & \targ{}  & \gate{R_x(2\beta)} & \gate{R_{CD}(\alpha)} & \meter{}
\end{quantikz}
\end{center}    
\begin{center}
\begin{quantikz}
\lstick{$\ket{0}$} & \gate{H} & \ctrl{2} & \qw       & \qw   & \qw       & \ctrl{2} & \gate{R_x(2\beta)} & \gate{R_{CD}(\alpha)} & \meter{} \\
\lstick{$\ket{0}$} & \gate{H} & \qw      & \ctrl{1}  & \qw   & \ctrl{1}  & \qw      & \gate{R_x(2\beta)} & \gate{R_{CD}(\alpha)} & \meter{} \\
\lstick{$\ket{0}$} & \gate{H} & \targ{}  & \targ{}   & \gate{R_z(2\gamma)} & \targ{}   & \targ{}  & \gate{R_x(2\beta)} & \gate{R_{CD}(\alpha)} & \meter{}
\end{quantikz}
\end{center}    
\end{figure*}

{\bf Optimization of Clusters:} 
A significant hurdle in deploying variational quantum algorithms to large-scale hardware engineering problems is algorithmic scalability, as directly solving for the global cost Hamiltonian $H^{\mathtt{C1}}$ across an entire processor lattice leads to an exponential expansion of the state space. This leaves the optimization loop highly susceptible to classical optimization stagnation, excessive circuit depths, and barren plateaus. To ensure our framework remains computationally scalable for next-generation hardware, we introduce a localized "Cluster Stitching" protocol that decomposes the high-density lattice into overlapping\cite{Xu2026SurfaceCodeHamiltonian,Kandala2017HardwareEfficientVQE}, localized k-qubit clusters $C_n$ (e.g., k=4). The optimization process is then fully parallelized using a hybrid quantum-classical loop (e.g., the Adam optimizer), where the expectation value of each sub-Hamiltonian is minimized independently according to:
\begin{equation}
	\underset{\gamma,\beta,\alpha}{\rm{Inf}} \expval{\psi}{H^{\mathtt{Cn}}} = E_{\rm{floor}}^{\mathtt{Cn}}
\end{equation}

Because adjacent clusters share overlapping boundary qubits, the local constraints at the borders naturally align and satisfy one another during the parallelized sub-routines. Once the individual localized loops reach convergence, the resulting discrete assignments are "stitched" together into a unified, globally coherent frequency map ${(f_1,f_2,...,f_N)}$.  Validated through benchmarking on 8-qubit ladders and 16-qubit grids, this division of labor successfully transforms an otherwise intractable global search into a localized task that scales linearly with processor size, enabling automated, real-time calibration of large-scale noisy intermediate-scale quantum~(NISQ) hardware while maintaining a stable residual crosstalk.

\section{Results}
\label{sec:results}

Before presenting the DC-QAOA benchmark, we first establish a rigorous
classical baseline. We implement simulated annealing (SA) directly over
the same discretized cost landscape used by the quantum solver, i.e., SA optimizes the exact classical evaluation of $H^{\mathtt{C1}}$ for each
candidate frequency assignment, with no approximation beyond the
discretization described in Sec.~\ref{sec:discretization}. This ensures the comparisons that follow (Sec.~\ref{sec:benchmark}) are on a similar scale and every solver is scored against the identical cost function.

\subsection{Classical Baseline: Simulated Annealing}
\label{sec:sa_baseline}

\emph{Algorithm:} At each iteration, a single qubit $i$ is selected uniformly at random and proposed a new frequency assignment $f_i' \in \mathbb{F}$ drawn uniformly from the candidate set in Eq. (\ref{eq.FM}). The proposed move is accepted according to the standard Metropolis criterion,
\begin{equation}
    P_{\text{accept}} = \min\!\left(1,\ \exp\!\left(-\frac{\Delta H^{\mathtt{C1}}}{T}\right)\right),
    \label{eq:sa_accept}
\end{equation}
where $\Delta H^{\mathtt{C1}} = H^{\mathtt{C1}}(f') - H^{\mathtt{C1}}(f)$ is the change in cost induced by the
proposed move. Moves that lower the cost are always accepted moves that
raise it are accepted with a temperature-dependent probability, allowing
the search to escape local minima early on, while converging to a strictly
descending trajectory as $T\to0$.

The temperature follows a geometric cooling schedule,
\begin{equation}
    T_k = T_0 \left(\frac{T_{\text{end}}}{T_0}\right)^{k/n},
    \qquad k = 0, 1, \dots, n,
    \label{eq:sa_cooling}
\end{equation}
where $n$ is the total number of iterations and $T_0$, $T_{\text{end}}$
These are the initial and final temperatures. The best assignment encountered over the full run is retained independently of the annealing trajectory's final state, since late-stage thermal fluctuations can move the search away from a good solution it has already visited.

\emph{Hyperparameter tuning:} The cooling schedule parameters were tuned empirically against brute-force-verified optima on small test clusters before being applied to the full benchmark, rather than chosen arbitrarily: we swept $n$ from $2{,}000$ to $20{,}000$ iterations and tracked the fraction of random seeds that recovered the exact known optimum (Table~\ref{tab:sa_tuning}). Convergence was unreliable below $n \approx10{,}000$ on the tested cluster size; we adopt $n=20{,}000$, at which $30/30$ independent seeds recover the exact optimum with zero variance, as the classical baseline used throughout the remainder of Sec .~\ref{sec:results}.

\begin{table}[h]
\centering
\begin{tabular}{lccc}
\hline
$n$ (iterations) & $T_0$ & $T_{\text{end}}$ & Seeds at optimum \\
\hline
2{,}000  & 100.0 & 0.001 & 19/30 \\
5{,}000  & 100.0 & 0.001 & 26/30 \\
10{,}000 & 100.0 & 0.001 & 29/30 \\
20{,}000 & 100.0 & 0.001 & 30/30 \\
\hline
\end{tabular}
\caption{Empirical convergence of SA against the brute-force optimum,
per iteration budget, on the reference 4-qubit cluster used throughout
Sec.~\ref{sec:results}. Reported over 30 independent random seeds.}
\label{tab:sa_tuning}
\end{table}

\emph{Reproducibility and fairness:}Performance of SAs is sensitive to under-tuning. An insufficiently annealed run can substantially under-perform its true capability, giving a misleadingly weak classical baseline. We report that this tuning process is explicit rather than presenting only the final,
best-tuned configuration. All solver comparisons in this section use matched evaluation budgets across SA and the quantum solvers wherever directly comparable, and we report mean and standard deviation over independent seeds for every solver rather than a single run.

\begin{algorithm}[H]
\caption{Simulated annealing over the discretized cost landscape}
\label{alg:sa}
\begin{algorithmic}
\Require candidate frequency set $F$, cost function $H^{\mathtt{C1}}$, iterations $N$, $T_0$, $T_{\text{end}}$
\State Initialize $f_i \leftarrow$ random element of $F$ for each qubit $i$
\State $f^{\text{best}} \leftarrow f$; \quad $H^{\text{best}} \leftarrow H^{\mathtt{C1}}(f)$
\For{$k = 0$ to $n-1$}
    \State $T \leftarrow T_0 (T_{\text{end}}/T_0)^{k/n}$
    \State Select qubit $i$ uniformly at random
    \State Propose $f_i' \leftarrow$ random element of $F$
    \State $\Delta H^{\mathtt{C1}} \leftarrow H^{\mathtt{C1}}(f \text{ with } f_i \to f_i') - H^{\mathtt{C1}}(f)$
    \If{$\Delta H^{\mathtt{C1}} \le 0$ \textbf{or} $\text{rand}() < \exp(-\Delta H^{\mathtt{C1}} / T)$}
        \State $f_i \leftarrow f_i'$
        \If{$H^{\mathtt{C1}}(f) < H^{\text{best}}$}
            \State $f^{\text{best}} \leftarrow f$; \quad $H^{\text{best}} \leftarrow H^{\mathtt{C1}}(f)$
        \EndIf
    \EndIf
\EndFor
\State \Return $f^{\text{best}}$
\end{algorithmic}
\end{algorithm}
The effectiveness of the DC-QAOA framework is evidenced by the successful navigation of the complex spectral landscape and the subsequent convergence to a hardware-optimized frequency map. As illustrated in Fig. \ref{fig:big_wide_fig}, the cost Hamiltonian exhibits a non-convex topography, with divergent energy barriers that represent physical "hard constraints," such as neighbor collisions and gate-resonance clashes. These high-penalty regions act as impenetrable boundaries within the variational loop, ensuring that any viable solution strictly avoids leakage-prone or crosstalk-heavy frequency configurations. The darker "Goldilocks Valleys" in the landscape denote the targeted spectral windows where qubit-qubit interactions are optimized for high-fidelity cross-resonance operations while maintaining maximum isolation from unwanted spectator effects.

\subsection{Benchmark: Solver Comparison}
\label{sec:benchmark}
\emph{Numerical Convergence and Optimization Dynamics:} The optimization dynamics and convergence behavior of the DC-QAOA framework were benchmarked across two distinct processor geometries: an 8-qubit ladder manifold and a high-density 16-qubit grid. Initialized in a purely stochastic state characterized by high-density spectral overlaps and severe constraint violations, the hybrid quantum-classical loop demonstrated a rapid and robust descent toward the target global minima. The distinctive "jagged" trajectory observed in the early iterations of the loss curves reflects the optimizers successful navigation through the non-convex landscape's steep penalty ridges.
Unlike standard variational mixers that regularly stall when encountering these divergent energy walls, the digitized counterdiabatic steering $H_{CD}$ provided the necessary quantum gradients to bypass non-physical local minima. Across both topologies, the system successfully converged to a stable plateau at a total residual crosstalk floor. This floor represents the physical limit of spectral isolation for the system, the mathematically unavoidable ZZ crosstalk dictated by the hardware's fixed coupling strengths and transmon anharmonicities.

\emph{8-Qubit Ladder:} The frequency mapping achieved for the 8-qubit ladder configuration  (decomposed into overlapping 4-qubit clusters, $C_0$ through $C_3$) demonstrates the efficacy of the framework under tight local constraints. The finalized allocation map successfully "quenched" all primary $\mathtt{A1}$
and $\mathtt{C1}$ collisions, positioning the assigned qubit frequencies entirely within the localized, low-crosstalk "Goldilocks Valleys." Crucially, the successful execution of the cluster-stitching protocol is validated by the uniform convergence of the overlapping sub-graphs. By incorporating the three-body $\mathtt{S1}$ spectator penalties, the algorithm strategically managed the detuning of the inner qubits relative to their nearest neighbors. This effectively shielded the interior nodes of the ladder from multi-body interference while maintaining a detuning profile that preserves viable operating speeds for the cross-resonance gates.
\begin{figure}[htbp]
    \centering
    \includegraphics[width=1.0\linewidth]{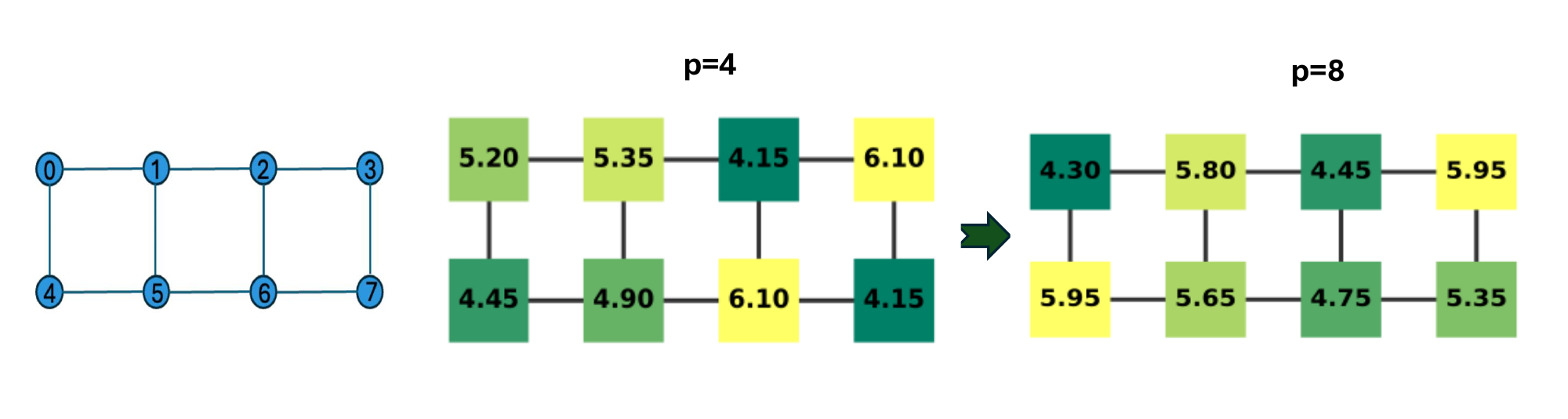}
    \caption{Mapping of the multi-tiered penalty system used in the variational optimization loop for the 8-qubit ladder system. The landscape highlights the divergent energy barriers that the DC-QAOA must navigate. Optimal frequency assignments are identified within the localized minima.}
    \label{fig:placeholder}
\end{figure}

\emph{Scalability towards 16 Qubits:}  To evaluate the scalability of the framework in higher-connectivity regimes, we deployed DC-QAOA on a 16-qubit square grid, decomposed into four overlapping clusters ($C_0$ through $C_3$) via the cluster-stitching protocol. In this two-dimensional architecture, the density of forbidden spectral regions is substantially higher than in the ladder case, since interior qubits must simultaneously satisfy pairwise constraints with up to four neighbors. The cluster-stitching protocol reconciled shared boundaries between adjacent clusters without requiring an increase in variational circuit depth or per-cluster optimization budget as the lattice size grew, consistent with the linear scaling. Across both the 8-qubit ladder and 16-qubit grid, we observe that increasing the circuit depth from $p=4$ to $p=8$ reduces the residual per-cluster cost, as expected from the counterdiabatic ansatz convergence behavior. We also note a structural feature of the resulting assignments: diagonally-positioned qubits that share a common neighbor (i.e., second-nearest-neighbor spectator pairs crossing cluster boundaries) occasionally receive similar frequencies. This is a direct consequence of the greedy cluster-stitching order of cross-cluster spectator constraints, which are not seen by any single local optimizer, and highlights that strict global spectator isolation in two-dimensional topologies may require either larger overlapping clusters or an iterative stitching pass. We treat this as a direction for future work rather than a limitation of DC-QAOA itself.

\begin{figure}[htbp]
    \centering
    \includegraphics[width=1.0\columnwidth]{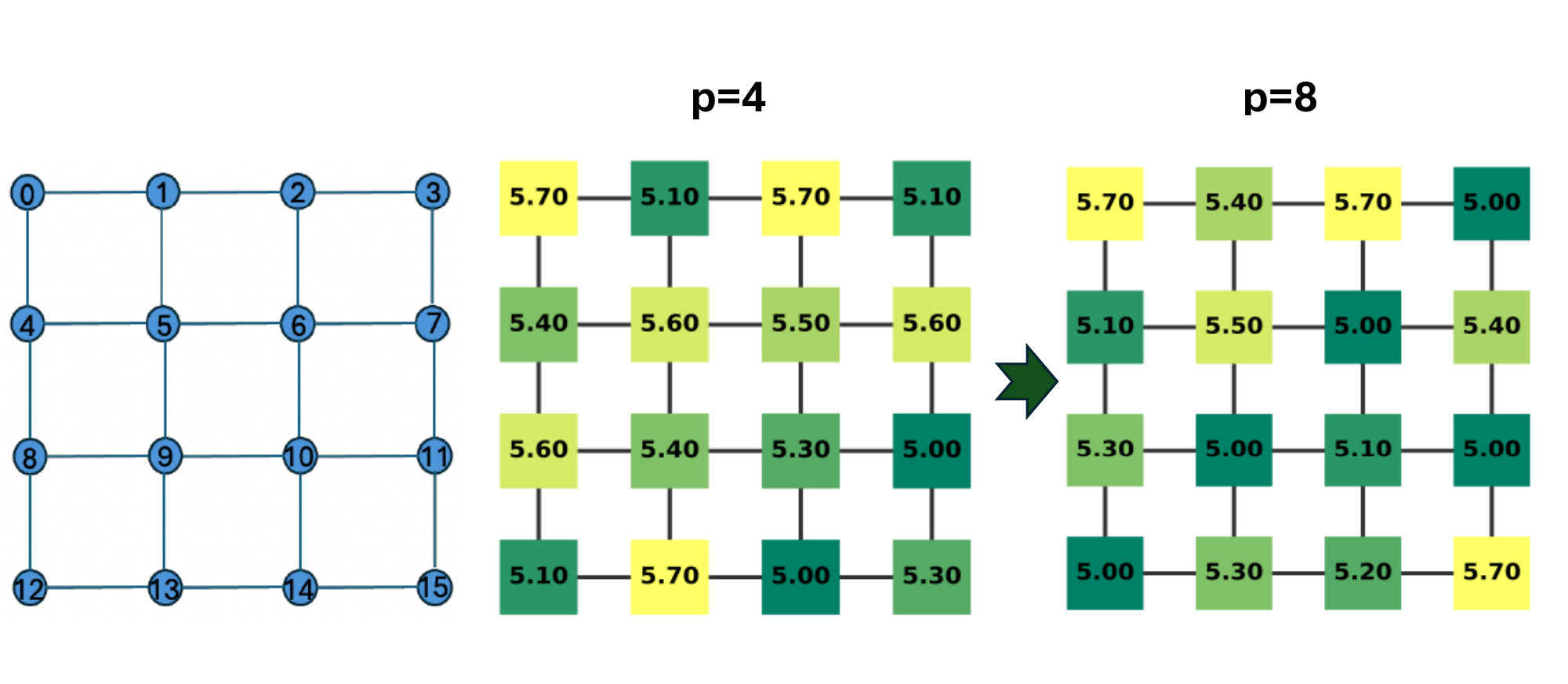}
     \caption{ Frequency allocation on a 16-qubit square lattice obtained with the DC-QAOA framework, shown for circuit depths $p=4$ (left) and $p=8$ (right). Each node color represents the frequency assigned to a qubit by the variational optimization loop acting on the penalty Hamiltonian $H_C$. Some spectator qubits in diagonally-opposite positions receive similar frequencies; this residual degeneracy arises because the local cluster-stitching protocol does not enforce constraints between qubits in non-adjacent clusters, and can be mitigated by extending the constraint graph to include all-to-all spectator interactions or by tightening the per-constraint penalty cutoff. }
    \label{fig:square1}
\end{figure}
DC-QAOA was deployed on the 8-qubit ladder, 13-qubit hexagonal and the 16-qubit square grid via the cluster-stitching protocol, decomposing the 16-qubit lattice into four overlapping clusters ($C_0$ through $C_3$) and producing globally-consistent frequency assignments across all tested clusters at both scales. As in the 8-qubit case, local constraints at shared cluster boundaries were reconciled automatically during stitching, without requiring an increase in circuit depth or per-cluster optimization budget as the lattice size grew.
\begin{figure}[htbp]
    \centering
    \includegraphics[width=1.0\columnwidth]{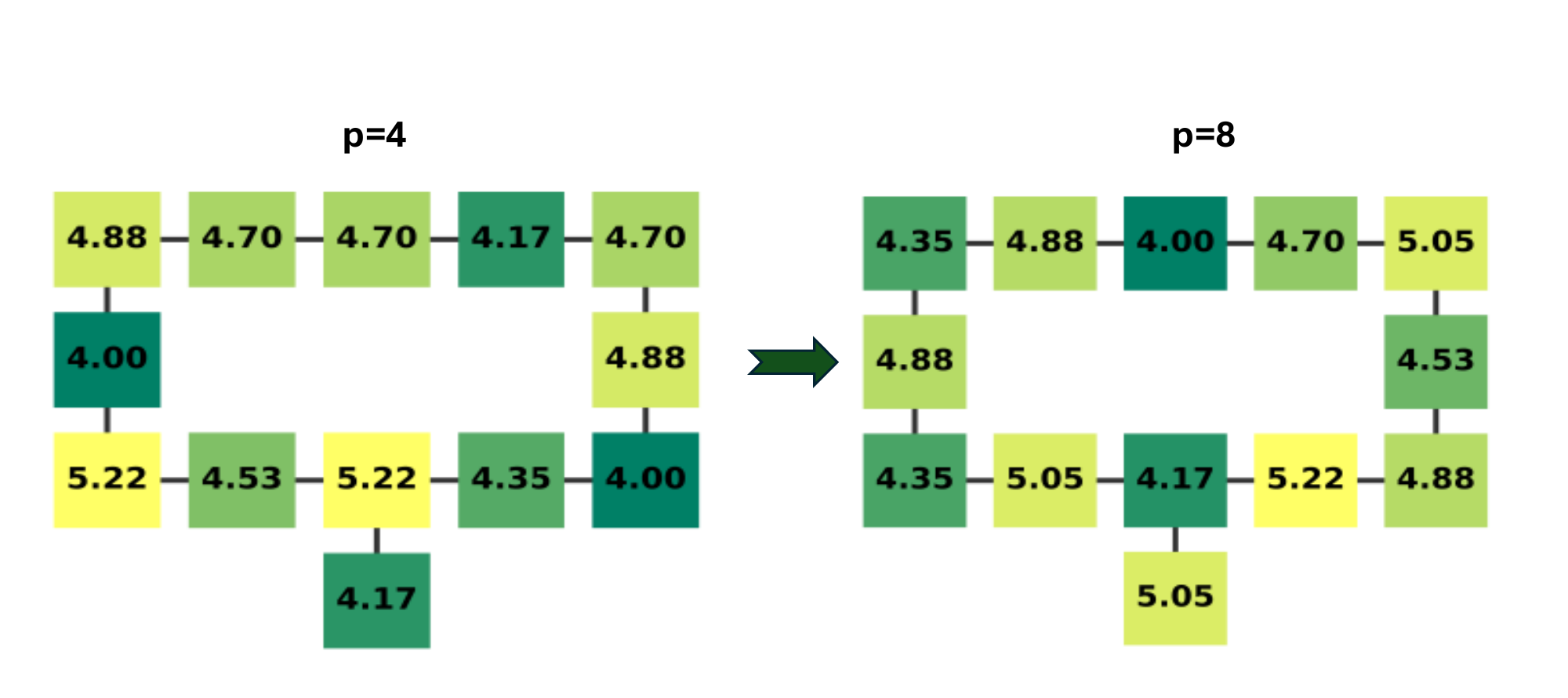}
     \caption{ Frequency allocation on a 13-qubit hexagonal ring obtained with the DC-QAOA framework, acting on the multi-tiered penalty Hamiltonian $H_C$. Node colors indicate the frequencies assigned by the variational optimization loop, which navigates the divergent energy barriers associated with the $\mathtt{A1}$, $\mathtt{C1}$, and $\mathtt{S1}$ collision mechanisms to reach the localized minima corresponding to collision-free assignments.}
    \label{fig:ibm}
\end{figure}
Relative to the standard transverse-field mixer alone, incorporating the digitized counterdiabatic term improved the recovered cost in the majority of tested clusters, consistent with its intended role of correlating the search direction with the local collision-penalty structure rather than exploring the state space independently of it. An ablation study directly comparing the correlated two-qubit counterdiabatic structure ($\hat{Y}_i\hat{Z}_j+\hat{Z}_i\hat{Y}_j$) against a simplified single-qubit alternative confirms this correlation is a meaningful contributor to the observed improvement, rather than an incidental effect of the additional variational parameters.

A few limitations of the present implementation are worth noting, each pointing to a concrete direction for extending the framework. The counterdiabatic improvement reported above is measured at the per-cluster level, before stitching, and does not by itself guarantee a globally collision-free assignment once clusters are reconciled; in the 16-qubit square-lattice case, pairs of qubits sitting in non-adjacent clusters, notably second-nearest-neighbor spectator pairs along the diagonal, are not seen by any single local optimizer, producing a residual degeneracy in the final map that is more pronounced as the lattice connectivity increases. These are intrinsic to the one-shot, locally-scoped nature of the current stitching protocol rather than to DC-QAOA itself, and admit several natural remedies: enlarging the cluster window to absorb a wider neighborhood of constraints, iterating the stitching order so that earlier clusters can be revisited once downstream boundary values are known, or promoting the cost Hamiltonian to explicitly include second-nearest-neighbor spectator terms so that the optimizer sees these constraints directly. Each of these is a straightforward extension of the framework presented here, and we expect that combining them will enable fully collision-free allocation at larger lattice sizes without altering the physics-derived Hamiltonian or the cluster-stitching principle thaunderpins in this work.
\section{Discussion}
In quantum architectures, where qubits are flux tunable, such as fluxonium or asymmetric transmon, the best operating points lie in the vicinity of the sweet spot, in which coherence time is maximum. This puts a limitation on the tunability of the quantum frequency. Therefore, in this work, we do not rely on this flux tunability, and we consider the qubit to have nearly fixed frequency during the entire optimization, which leaves room for optimization. A naive frequency-allocation strategy enforces only coarse spectral separation, treating any sufficiently large detuning as safe. This criterion can be misleading in fixed-frequency architectures because the residual ZZ interaction does not decrease monotonically with detuning. Consequently, a frequency map optimized solely to avoid neighboring-qubit degeneracy can place coupled qubits inside this anharmonicity-driven resonance region. The resulting residual parasitic interaction can be catastrophic despite an apparently conservative detuning margin. 

The standard approach to currently practice design process is assigning frequency to qubits and applying method to parasitic interactions such as dynamical decoupling, echoed pulse sequences, or active coupler biasing~\cite{Yan2018TunableCoupler,Mundada2019}. The problem with these methods is that processor absorbs base error at ideal state. Due to the fact that parasitic interactions at the ideal state are persistent.~We propose combining error minimization and frequency allocations, in which before allocating frequency in different cells we minimize the static erroneous error base collection, which guarantees to minimize the phase error at the ideal state. Repeated reset-and-re-initialization cycles are time-consuming and are themselves driven by the need to clear accumulated phase error from always-on parasitic interactions. The circuit we propose addresses this at its source: by minimizing the static phase-accumulation terms in the Hamiltonian before frequency assignment, the framework reduces the error budget that would otherwise require periodic resets to suppress.

Explicitly incorporating two-qubit and three-qubit interactions into the allocation objective allows the optimizer to identify assignments that are both collision-free and resonance-free by construction. This addresses parasitic interactions during allocation; it thereby eliminates a class of hardware failure that purely combinatorial or degeneracy-based allocation schemes cannot detect. This distinction has direct consequences for scalability and gate scheduling.~Suppressing ZZ crosstalk through frequency allocation reduces reliance on real-time cancellation and active mitigation hardware, such as tunable couplers.~It also reduces the associated calibration overhead and control-line complexity.~Static ZZ interactions accumulate phase continuously during idle periods, whether or not a gate is executed.~Addressing them during allocation is therefore particularly advantageous for parallel gate execution. Qubits whose frequencies are jointly optimized against $H^{\mathtt{C1}}$ can remain idle alongside active neighboring gates without contributing uncontrolled coherent phase errors. Post-hoc correction schemes, by contrast, typically require scheduling awareness or duty cycled suppression, limiting achievable parallelism. Including the three-body interaction alongside the pairwise contribution also captures spectator-induced gate degradation that pairwise-only or ZZ-avoidance strategies would miss. A target control pair can satisfy every pairwise constraint yet still suffer reduced gate fidelity because of a spectator qubit. This effect becomes increasingly important as processor size and connectivity grow.

We developed a hardware-aware variational framework for frequency allocation in superconducting quantum processors. The framework encodes physical crosstalk mechanisms, including $\mathtt{A1}$ collisions, $\mathtt{C1}$ clashes, and $\mathtt{S1}$ spectator effects, in a discrete cost Hamiltonian. We used DC-QAOA to address the resulting non-convex optimization problem.~To improve scalability, we introduced a cluster-stitching protocol that constructs global frequency maps by satisfying local constraints. We demonstrated the approach on an eight-qubit ladder and established its applicability to both classical and quantum optimizers.~The current classical simulated annealing implementation outperformed the quantum optimizers, while cluster stitching remained independent of the optimization method. YZ-type counterdiabatic steering terms also improved state-vector stability near divergent penalty terms, where standard mixers tend to stall.~Future work will extend the protocol to higher-connectivity architectures, including heavy-hex and square lattices, and to tunable-qubit systems. Further advances in quantum optimization will be needed to assess the quantum potential advantage.~The framework provides a foundation for automated software hardware co-design and dynamic frequency calibration in scalable superconducting quantum processors.

\begin{acknowledgments}
	This research received funding from the Horizon Europe OpenSuperQPlus100 project (Grant Agreement No. 101113946).  
\end{acknowledgments}

\bibliography{menuscriptNotes}

\appendix
\section{Types of collision}
\label{app.A}
\begin{center}
\begin{small}
\begin{table}[h]
\begin{tabular}{c c c c c} 
 \hline
 & & Definition & Participants & Bounds \\ [0.5ex] 
 \hline\hline
 Addressability &$\mathtt{A1}$ & $|f_i-f_j|\geq\delta^{\mathtt{A1}}$& $(i,j)\in \mathbb{E}$ & 17MHz\\ 
  & A2 & $|f_i-f_j-\alpha_j|\geq\delta_{A2}$& $(i,j)\in \mathbb{E}$& 30MHz \\
 Entanglement&C1& $f_i+\alpha_i \leq f_de \leq f_i$&$(i,j)\in \mathbb{E}$&-\\
   & E1&$|f_d-f_i|\geq \delta_{E1}$&$(i,j)\in \mathbb{E}$& 17MHz\\
    &E2& $|f_d-f_i-\alpha_i|\geq \delta_{E2}$& $(i,j)\in \mathbb{E}$ & 30MHz\\
Spectators&S1&$|f_d-f_k|\geq \delta_{S1}$& $(i,j,k)\in N $& 17MHz\\
&S2& $|f_d-f_k-\alpha_k| \geq \delta_{S_2}$\\
\hline
\hline
\end{tabular}
\caption{In the graph, with $N$ nodes and $\mathbb{E}$ edges,
the above table represents the different collisions.}
\label{sec:T1}
\end{table}
\end{small}
\end{center}
In this section, we present and compare the results obtained for different lattice structures. The analysis is carried out at a central frequency of $\omega = 5.5 GHz$ with a frequency spacing $\Delta = 0.5~GHz$. For each lattice configuration, the energy distribution is calculated to investigate how the structural arrangement influences the system’s overall behavior. By examining various lattice geometries, we aim to identify the effects of connectivity, symmetry, and dimensionality on the resulting energy profiles and to understand how these factors contribute to the observed differences in physical properties.
\subsection{Correspondence with the main-text collision model}
\label{app:collision-correspondence}
Table~\ref{sec:T1} lists the complete set of collision categories relevant to fixed-frequency transmon architectures, following the Addressability, entanglement, and spectator classification of Ref.~\cite{Zhang2025FrequencyAllocation}. The cost Hamiltonian $H^{\mathtt{C1}}$ used in Sec.\ref{sec:method}, however, retains only three of these seven categories: $\mathtt{A1}$, $\mathtt{S1}$, and the drive-leakage condition we refer to in the main text as $\mathtt{C1}$.

We note explicitly that the main-text $H^{\mathtt{C1}}$, defined by the drive-frequency condition $\omega_{\rm drive}=\omega_j$ approaching the $\ket{1}\rightarrow\ket{2}$ transition $\omega_i+\eta_i$ of a neighboring transmon, corresponds to the $\mathtt{E2}$ row of Table~\ref{sec:T1} rather than the table's own $\mathtt{C1}$ row (which instead specifies the target operating window $f_i+\alpha_i\leq f_{d}\leq f_i$ for a well-tuned Cross-Resonance drive). To avoid ambiguity, all instances of `$\mathtt{C1}$'' in the main text should be read as referring to this $\mathtt{E2}$-type leakage condition; we retain the label $\mathtt{C1}$ there for consistency with the gate-clash terminology used throughout the algorithmic discussion, but the reader should identify it with Table~\ref{sec:T1}'s $\mathtt{E2}$ bound ($\delta_{E2}=30$~MHz) rather than its $\mathtt{C1}$ entry.

The remaining four categories, $\mathtt{A2}$, the table's own $\mathtt{C1}$ window, $\mathtt{E1}$, and $\mathtt{S2}$  are neglected in $H^{\mathtt{C1}}$ for the following reasons. $\mathtt{A2}$ and $\mathtt{E1}$ share the same functional form as $\mathtt{A1}$ and $\mathtt{E2}$ ,respectively, differing only in which transmon anharmonicity $\alpha$ enters the collision condition; at the frequency spacing $\Delta=0.5$~GHz used throughout this work, both are subdominant to their $\mathtt{A1}/\mathtt{E2}$ counterparts and do not independently bind the optimization. The table's $\mathtt{C1}$ window is a design constraint on drive placement rather than a pairwise collision between fixed qubit frequencies, and is instead enforced implicitly through the $ H^{\mathtt{Tg}}$ gate-time balancing term of Eq.~\eqref{eq.HTg}. $\mathtt{S2}$ introduces a second-order spectator correction with bound $\delta_{S2}$ comparable to $\delta_{S1}$, we neglect it at the cluster sizes and consider that it follows immediately from the same three-body construction as $H^{\mathtt{S1}}$ and would enter $H^{\mathtt{C1}}$ as an additional term of the form
\begin{equation}
    H_{S2} = \sum_{(i,j,k)\in T} \chi_{ijk}^{(S2)}\,\Theta\!\left(\delta_{S2} - |f_j - f_k - \alpha_k|\right).
\end{equation}

\section{Derivation of the Digitized Counterdiabatic Mixer}
\label{app.B}

This section derives the digitized counterdiabatic mixer $H_{CD}$ used in the main text directly from the cost Hamiltonian $H^{\mathtt{C1}}$, via the nested-commutator (NC) approximation to the adiabatic gauge potential~\cite{Chandarana2022DCQAOA,delCampo2013ShortcutsAdiabaticity}.

\subsection{Nested-commutator ansatz}
For a parameterized Hamiltonian interpolating between a mixer $H^{\mathtt{M}}$ and a target cost Hamiltonian $H^{\mathtt{C1}}$, the exact adiabatic gauge potential $\mathcal{A}_\lambda$ satisfies $[\partial_\lambda H, \mathcal{A}_\lambda] + i\partial_\lambda H = 0$ up to terms that vanish on the instantaneous eigenbasis. Because $\mathcal{A}_\lambda$ is generally intractable for many-body $H^{\mathtt{C1}}$, we adopt the first-order NC ansatz,
\begin{equation}
    H_{CD} = \mathcal{A}_\lambda^{(1)} = i\,\alpha_p\,[H^{\mathtt{M}}, H^{\mathtt{C1}}],
    \label{eq:NC1}
\end{equation}
with $H^{\mathtt{M}} = \sum_i \hat{X}_i$ the standard transverse mixer and $\alpha_p$ the variational counterdiabatic weight at layer $p$.

\subsection{Cost Hamiltonian decomposition}
The full cost Hamiltonian separates into single-body, two-body, and three-body sectors,
\begin{equation}
    H^\mathtt{C1} = \underbrace{\sum_k P(f_k)\hat{Z}_k}_{H^\mathtt{C1}(1)} \;+\; \underbrace{\sum_{i<j} \mathcal{W}_{ij}\hat{Z}_i\hat{Z}_j}_{H^{\mathtt{C1}}(2)} \;+\; \underbrace{\sum_{i<j<k} \chi_{ijk}\hat{Z}_i\hat{Z}_j\hat{Z}_k}_{H^\mathtt{C1}{(3)}}.
    \label{eq:HCdecomp}
\end{equation}
Since the commutator in Eq.~\eqref{eq:NC1} is linear, $H_{CD}$ decomposes accordingly, and each sector can be evaluated separately using the Pauli identity
\begin{equation}
    [\hat{X}_i, \hat{Z}_i] = -2i\,\hat{Y}_i.
    \label{eq:pauli}
\end{equation}

\subsection{Two-body contribution}
For a single term $\mathcal{W}_{ij}\hat{Z}_i\hat{Z}_j$ with $i<j$, only $\hat{X}_i$ and $\hat{X}_j$ from $H^{\mathtt{M}}$ fail to commute:
\begin{align}
    [\hat{X}_i,\, \mathcal{W}_{ij}\hat{Z}_i\hat{Z}_j] &=& \mathcal{W}_{ij}\,[\hat{X}_i,\hat{Z}_i]\,\hat{Z}_j = -2i\,\mathcal{W}_{ij}\,\hat{Y}_i\hat{Z}_j, \\
    [\hat{X}_j,\, \mathcal{W}_{ij}\hat{Z}_i\hat{Z}_j] &=& \mathcal{W}_{ij}\,\hat{Z}_i\,[\hat{X}_j,\hat{Z}_j] = -2i\,\mathcal{W}_{ij}\,\hat{Z}_i\hat{Y}_j. 
\end{align}
Summing over all pairs and substituting into Eq.~\eqref{eq:NC1},
\begin{equation}
    H_{CD}^{(2)} = 2\alpha_p \sum_{i<j} \mathcal{W}_{ij}\left(\hat{Y}_i\hat{Z}_j + \hat{Z}_i\hat{Y}_j\right).
    \label{eq:HCD2}
\end{equation}
This reproduces the $YZ$-type mixer used in the main text, with coupling coefficients
\begin{equation}
    A_{ij} = 2\alpha_p\, \mathcal{W}_{ij},
    \label{eq:Aij}
\end{equation}
showing explicitly that the counterdiabatic coefficients are proportional to the same $\mathtt{A1}/\mathtt{C1}$ collision-penalty weights that define $H^\mathtt{C1}{(2)}$.
\subsection{Three-body contribution}
The identical procedure applied to $H^\mathtt{C1}{(3)}$ gives, for each term $\chi_{ijk}\hat{Z}_i\hat{Z}_j\hat{Z}_k$ with $i<j<k$,
\begin{align}
    [\hat{X}_i + &\hat{X}_j + \hat{X}_k,\; \chi_{ijk}\hat{Z}_i\hat{Z}_j\hat{Z}_k] \\&
    = -2i\,\chi_{ijk}\left(\hat{Y}_i\hat{Z}_j\hat{Z}_k + \hat{Z}_i\hat{Y}_j\hat{Z}_k + \hat{Z}_i\hat{Z}_j\hat{Y}_k\right),
\end{align}
so that
\begin{equation}
    H_{CD}^{(3)} = 2\alpha_p \sum_{i<j<k} \chi_{ijk}\left(\hat{Y}_i\hat{Z}_j\hat{Z}_k + \hat{Z}_i\hat{Y}_j\hat{Z}_k + \hat{Z}_i\hat{Z}_j\hat{Y}_k\right).
    \label{eq:HCD3}
\end{equation}
This three-body counterdiabatic term is sourced directly by the S1 spectator penalties $\chi_{ijk}$.

\subsection{Single-body contribution}
Finally, the single-body sector contributes
\begin{equation}
    H_{CD}^{(1)} = 2\alpha_p \sum_k P(f_k)\,\hat{Y}_k,
    \label{eq:HCD1}
\end{equation}
a locally biased single-qubit rotation sourced by the hardware-bandwidth penalty $P(f_k)$.

\subsection{Full ansatz and truncation}
The complete first-order NC counterdiabatic term is therefore
\begin{equation}
    H_{CD} = H_{CD}^{(1)} + H_{CD}^{(2)} + H_{CD}^{(3)}.
    \label{eq:HCDfull}
\end{equation}
In the implementation described in the main text, we retain only the two-body term $H_{CD}^{(2)}$ Eq.\eqref{eq:HCD2}, which captures the dominant counterdiabatic steering associated with the $\mathtt{A1}$ and $\mathtt{C1}$ hard constraints at fixed circuit depth. The single-body term $H_{CD}^{(1)}$ and three-body term $H_{CD}^{(3)}$ are neglected at this order; including $H_{CD}^{(3)}$ would compensate more directly for S1-induced ruggedness in the optimization landscape at the cost of additional two-qubit gates per layer, and is left for future work.
	
\end{document}